\documentclass[twocolumn,tightenlines,floatfix]{revtex4}

\usepackage{epsfig} 
\usepackage{amssymb} 
\usepackage{amsmath} 
\usepackage{amsfonts}

\newcommand{\be}{\begin{equation}}
\newcommand{\ee}{\end{equation}}

\newcommand{\gev}{{\rm GeV}}
\newcommand{\tev}{{\rm TeV}}

\def\lsim{\buildrel < \over {_{\sim}}}

\begin{document}

\title{High energy leptons from muons in transit}

\author{Alexander Bulmahn and Mary Hall Reno}
\affiliation{Department of Physics and Astronomy, University of Iowa, Iowa City, IA 52242}

\begin{abstract}

The differential energy distribution for electrons and taus produced from lepton pair production from  muons in transit through materials is numerically evaluated.  We use the differential cross section to calculate underground lepton fluxes from an incident atmospheric muon flux, considering contributions from both conventional and prompt fluxes.  An approximate form for the charged current differential neutrino cross section is provided and used to calculate single lepton production from atmospheric neutrinos. We compare the fluxes of underground leptons produced from incident muons with those produced from incident neutrinos and photons from muon bremsstrahlung.  We discuss their relevance for underground detectors. 
\end{abstract}

\maketitle

\section{Introduction}

Atmospheric muon fluxes at sea level have pushed neutrino experiments underground or under ice, however, even at a depth of more than 1.5 km under ice, the muon to muon neutrino induced muon ratio is on the order of $10^6$. Much of the work in detectors like AMANDA \cite{amanda} and IceCube \cite{icecube} has concentrated on
upward $\nu_\mu\to \mu$ conversions to avoid the downward
muon background. The new DeepCore module \cite{deepcore}, with a threshold of $\sim 10$ GeV and
situated within IceCube, is designed to allow some of the downward muon neutrino flux to be studied where the IceCube detector acts as a muon veto for the background to muons produced in the DeepCore detector.   The DeepCore module also offers the potential to measure $\nu_\tau\to \tau$ and $\nu_e\to e$ conversions \cite{deepcore}.

The downward atmospheric muon flux itself \cite{Lipari1,bartol,honda,Bugaev} presents an opportunity to explore the predictions of quantum electrodynamics. 
A well known process is charged lepton pair production
\cite{br,kelner,Kelner:1998mh,Kelner:2001fg,abs,kp}, e.g.,
$$\mu A\rightarrow \mu e^+ e^- X\ .$$
This process is the primary contributor to the muon energy loss, along
with bremsstrahlung, photonuclear and ionization interactions in transit through materials \cite{dedx}. Most of the energy loss in electron positron pair production occurs through
small changes in muon energy. Nevertheless,
occasionally the charged lepton produced carries a large energy. Given
that there are so many downward-going muons, one may have the opportunity through
rare high energy lepton production to explore physics at high energy scales.

Through charged lepton pair production, one has the potential to see
evidence for the onset of charm particle production of muons in the atmosphere \cite{prs,tig,enberg}.  At high energies, the long decay length of light mesons does not allow them to decay before the surface of the earth.  Studying charged lepton signals underground presents another opportunity for determining the contributions from charm production to the atmospheric muon flux.
  
Charged tau pair production is also of interest.
Atmospheric production of tau neutrinos is quite low, since tau production comes
from $D_s$ and $b$ meson production in cosmic ray interactions with the air nuclei
\cite{pr,stasto}. At high energies, neutrino oscillations for $\nu_\mu\rightarrow
\nu_\tau$ are also suppressed. Atmospheric muon production of tau pairs may produce a
tau signal (accompanied by a muon) or ultimately a tau neutrino flux in an
energy regime where few are expected. 

In Sec. II, we describe how our prior work for calculating the cross section for lepton pair production from muons scattering with atomic targets can be applied to numerical calculations of the energy distribution of the leptons produced for fixed muon energy \cite{br}. The approximate analytic form of Tannenbaum \cite{tannenbaum} can be translated to an electron energy
distribution which agrees well with our numerical results. The tau energy distribution must be evaluated numerically.
To compare the fluxes of underground leptons produced from incident atmospheric muons with those produced from incident neutrinos, we present an approximate form of the neutrino-nucleon differential cross section that agrees well with numerical evaluations for a large range of incident neutrino energies and energy transfers to the produced charged lepton.
We review in Sec. III the steps to go from a muon or neutrino flux to an electron or tau flux, and we describe the parameterizations for atmospheric muon and neutrino fluxes used here.  

Our calculation can be applied to many detector geometries.  In Sec. IV, we have focused our calculation of underground electron production for the large underground Cherenkov detector IceCube.  Because electromagnetic showers produced by electrons are difficult to distinguish from those produced by photons, we compare the flux high of energy electrons produced via pair production with that of high energy photons produced by muon bremsstrahlung.  In addition to electron signals, we also calculate the fluxes of underground taus produced by incident muons and tau neutrinos, and we explore the possibility of $\tau$ pair production signals at IceCube.  In addition to IceCube, we also explore the possibility of taus produced from incident muons and tau neutrinos in the mountains surrounding the High Altitude Water Cherenkov (HAWC) detector \cite{hawc}.  HAWC has the potential to see the subsequent tau decay showers.  In the final section, we discuss our results. 

\section{Production of charged leptons}

With our focus on lepton pair production by  muons in transit, we begin our discussion with our results for the electron or tau energy distribution for a fixed incident muon energy. 
Another source of electrons and taus, in this case single leptons, is from neutrino charged current interactions. We discuss this below.

\subsection{Lepton pair production}

We have presented in Ref. \cite{br} the formulas to evaluate the differential cross section for a charged muon to scatter with a nucleus of charge $Z$ and atomic number $A$. These formulas extend the work of Kel'ner \cite{kelner} and Akhundov et al. \cite{abs}
and others \cite{kp} so that they applicable to both
electron-positron pair production and tau pair production. We have not done the full calculation required for $\mu^+\mu^-$ production because
of the complication of identical particles in the final state.

\begin{figure}[h]
\includegraphics[width=2.4in,angle=270]{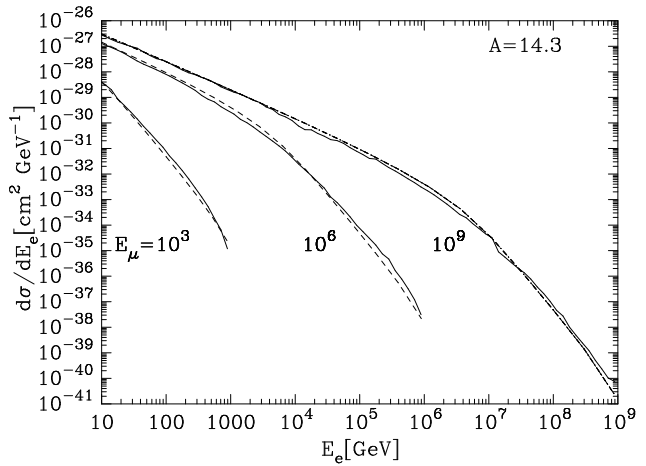}
\caption{Differential cross section as a function of electron energy
for $\mu A\rightarrow \mu e^+ e^- X$ for fixed muon energy.  Here $A=14.3$  and $Z= 7.23$ for water. The solid line shows our numerical result from Ref. \cite{br} and the dashed lines show the approximation of Tannenbaum \cite{tannenbaum} using
$v=(E_\mu-E_\mu')/E_\mu\simeq 2E_e/(E_\mu-E_e/2)$.}
\label{fig:dsdee}
\end{figure}

As we showed in Ref. \cite{br}, for electron positron pair production, the differential cross section is dominated by very low
momentum transfers to the target. Consequently, coherent scattering with the nucleus dominates the cross section, and the static nucleus approximation of Kel'ner \cite{kelner} is quite good.
In the numerical work below, we use the differential cross section as a function of $E_e$, the energy of the positron or electron. Our numerical results for the electron energy distribution for incident muon energies
of $10^3$ GeV, $10^6$ GeV and $10^9$ GeV are shown by the solid lines 
in Fig. \ref{fig:dsdee}. 

While we use our numerical results for the differential electron energy distribution, we note that Tannenbaum in Ref. \cite{tannenbaum}
has an approximate expression for the differential cross section as a function of $v=(E_\mu-E_\mu')/E_\mu$ for an
initial muon energy $E_\mu$ and final muon energy $E_\mu'$. In the limit of low momentum transfer to the nucleus,
$E_\mu v \simeq E_e+E_{\bar{e}}$, the sum of energies of the charged lepton pair. To first approximation,
$E_e\simeq vE_\mu/2$. In terms of $v$, Tannenbaum has \cite{tannenbaum}
\begin{eqnarray}
\nonumber
v\frac{d\sigma}{dv} &=& \frac{28}{9\pi}Z(Z+1)(\alpha r_l)^2\Bigl[(1+z^2)\ln(1+\frac{1}{z^2})-1\Bigr] \\ 
&& \times f(e,v).
\label{eq:tan}
\end{eqnarray}
Here $r_l$ is the classical radius of the lepton produced and $z=vm_\mu/4m_l$.  The function $f(e,v)$ can be expressed in two limiting regions as
\begin{equation}
\label{eq:fcases}
f(e,v) = \begin{cases}
\ln\Bigl(vE_\mu/6.67m_l\Bigr) & {\rm unscreened} \\
 \ln\Bigl(184.15m_l/m_eZ\Bigr) & {\rm fully\ screened\, .}
\end{cases}
\end{equation}
In comparing our numerical result to those found using Eq. (\ref{eq:tan}), we have taken the lower of the two values of $f(e,v)$. We have modified $v$ to access the high energy tail of the distribution, and with $v \simeq 2E_e/(E_\mu-\frac{1}{2}E_e)$, we can reproduce our numerical results as long as $v<4$.
This gives
\begin{equation}
\label{eq:ee}
E_e =\frac{2v}{4+v}E_\mu\ , \quad\quad
 v<4\frac{(1-m_\mu/E_\mu)}{(1+m_\mu/E_\mu)}\ ,
\end{equation}
neglecting $m_e$ compared to $m_\mu$.
The dashed lines
in Fig. \ref{fig:dsdee} show the differential cross section for
muon production of an electron of energy $E_e$ for $E_\mu=10^3\ \gev,\
10^6\ \gev$ and $10^9$ GeV with the approximation of Tannenbaum using
Eq. (\ref{eq:ee}). 

\begin{figure}[h]
\includegraphics[width=2.4in,angle=270]{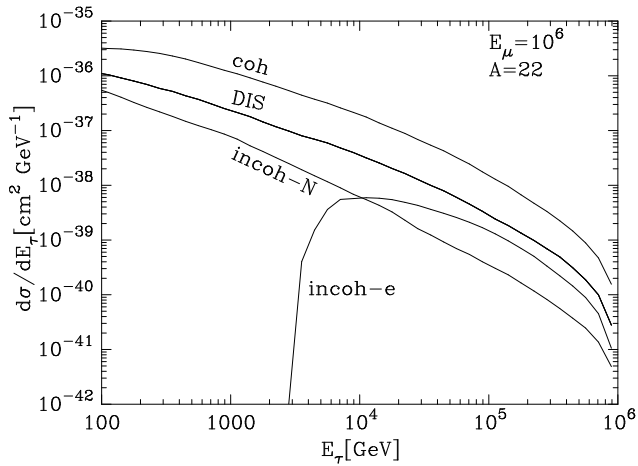}
\caption{Contributions to the differential cross section as a function of tau energy
for $\mu A\rightarrow \mu \tau^+ \tau^- X$ for fixed muon energy, $E_\mu = 10^6$ GeV.  Here $A=22$ and $Z=11$ is used for standard rock. Indicated are the coherent scattering contributions (coh), deep inelastic scattering (DIS) and scattering with individual nucleons (incoh-N) and electrons (incoh-e).}
\label{fig:dsdetau}
\end{figure}

The approximate form for the differential energy distribution of the produced electron does not have an easy correspondence to tau pair production \cite{br}. For tau pairs, there is a significant contribution from inelastic scattering of the muon with the target and higher momentum transfers need to be taken into account. For tau pairs, we can only use our numerical differential cross section.
Fig. \ref{fig:dsdetau} shows the scattering contributions to the differential distribution of the $\tau$ energy for $\tau^+\tau^-$ production by muons in transit through rock with an initial muon energy $E_\mu = 10^6 \ \gev$.

The contributions in Fig. \ref{fig:dsdetau} are the coherent scattering with the nucleus (coh), deep inelastic scattering (DIS), and scattering with individual nucleons (incoh-N)
and electrons (incoh-e). The curve showing incoherent scattering with atomic electrons demonstrates the threshold behavior for $\tau^+ \tau^- $ production.

In what follows, we use the numerical differential cross sections for $\ell = e$ or
$\ell = \tau$ production by $\mu\rightarrow \mu \ell^+\ell^- X$, where the incident muons are produced by cosmic ray interactions in the atmosphere.

\subsection{Neutrino charged current interactions}

Neutrino production of charged leptons is the primary focus of IceCube and other underground experiments. To compare charged lepton pair production by atmospheric muons to neutrino production of a single charged lepton in interactions with nucleons, it is helpful to have an approximate differential cross section for charged current neutrino interactions. 

In principle, the neutrino charged current cross section depends on the mass of the charged lepton produced.
At $E_\nu\simeq 100$ GeV, the $\nu_\tau N$ charged current cross section,
for isoscalar nucleon $N$, is about 80\% that for $\nu_\mu N$ \cite{kr1,kr2}. By $E_\nu= 1$ TeV, the ratio of the $\nu_\tau N$ to $\nu_\mu N$ charged current cross sections is about 0.95. We neglect the lepton mass corrections in what follows.

An approximate form for the charged current differential neutrino cross section at low energies can be found, e.g., in Ref. \cite{strumia},
\begin{eqnarray}
\label{eq:bardiff}
\frac{d\sigma_\nu}{dy} &=& \frac{2m_pG_f^2E_\nu}{\pi}
\Bigl(0.2+0.05(1-y)^2\Bigr)\\
\label{eq:bardiff2}
\frac{d\sigma_{\bar{\nu}}}{dy} &=& \frac{2m_pG_f^2E_\nu}{\pi}
\Bigl(0.05+0.2(1-y)^2\Bigr)
\end{eqnarray}
with inelasticity $y = 1-E_{\ell}/E_\nu$.  
While reasonable approximations to the neutrino nucleon differential
cross section at $E_\nu \sim 1$ TeV, 
Eqs. (\ref{eq:bardiff}) and (\ref{eq:bardiff2})
do not account for the $y$ dependence and energy dependence that come from increased contributions from sea quarks as neutrino
energies increase. They don't reflect
the violation of Bjorken scaling as the momentum transfer
increases. 

\begin{figure}[h]
\includegraphics[width=2.4in,angle=270]{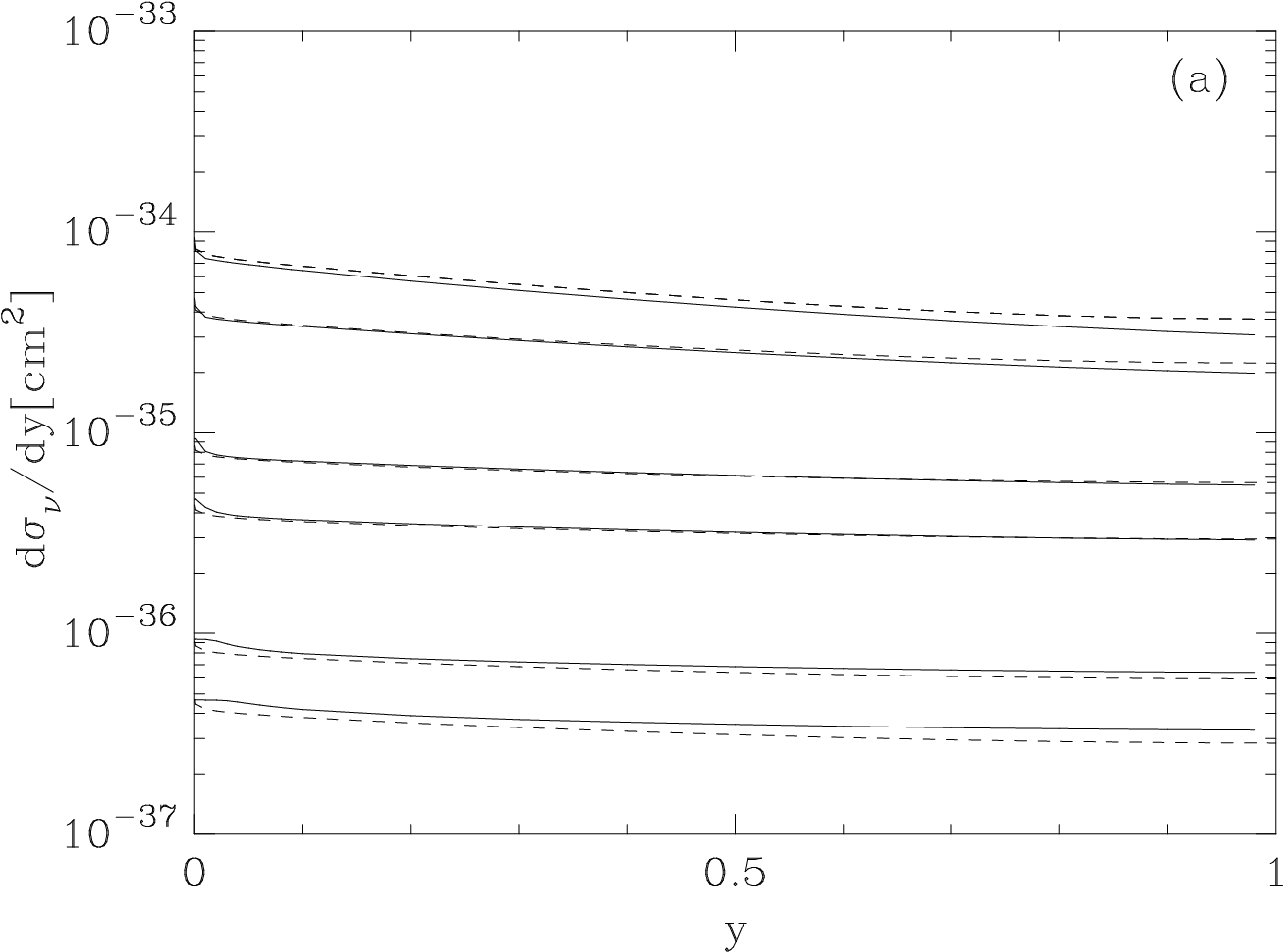}
\includegraphics[width=2.4in,angle=270]{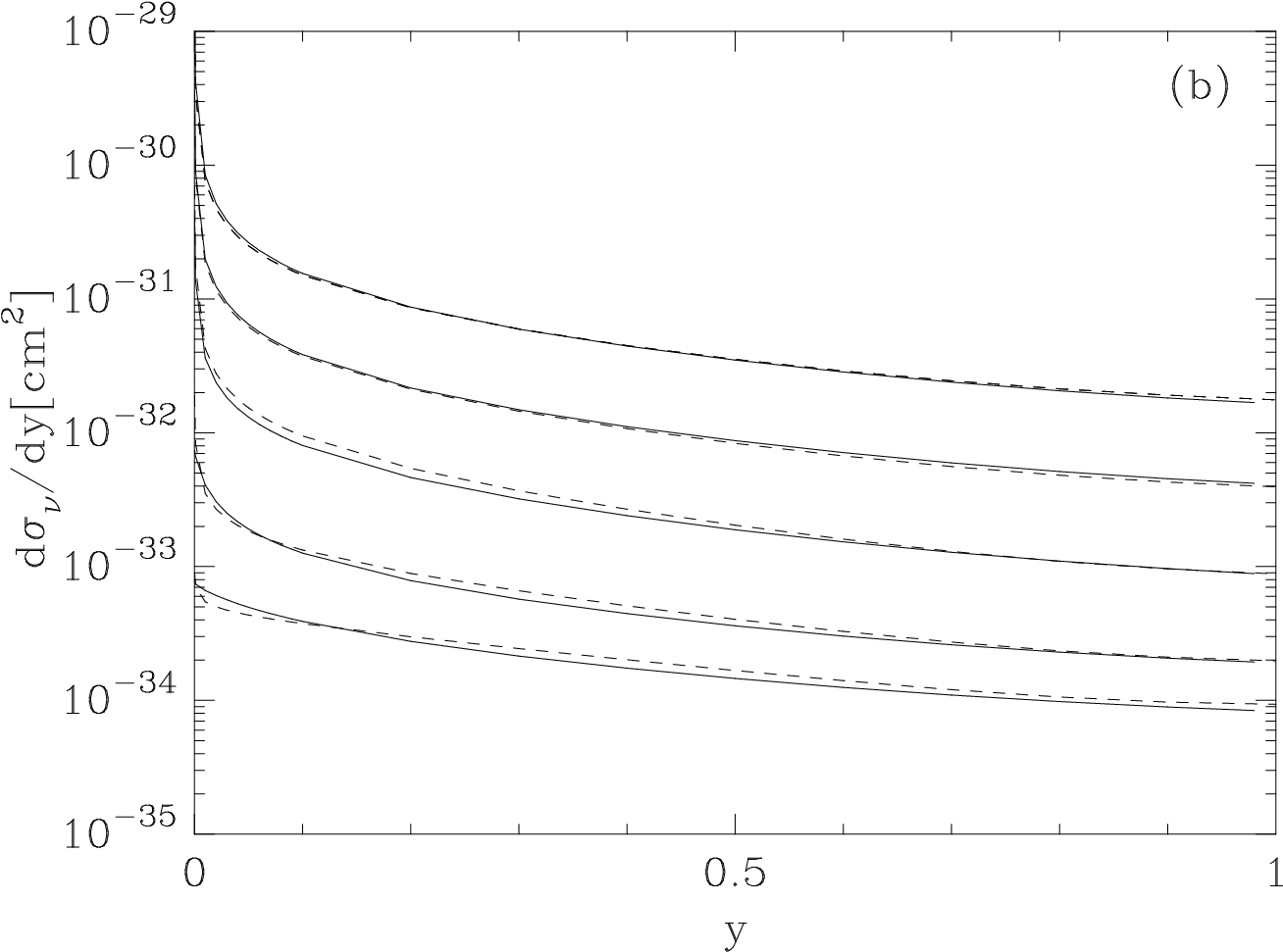}
\caption{Differential neutrino-nucleon cross section defined by Eq.
\ref{eq:brdiff} with parameters from Eqs. (\ref{eq:nudiffle}-\ref{eq:nudiffhe}). The solid lines represent numerical results using the CTEQ6 parton distribution functions \cite{cteq6} and the dashed lines are our approximate analytic formula.  Fig. (a) is the differential cross section for $E_\nu\leq 3.5\times 10^4$ with the fit parameters defined in Eq. (\ref{eq:nudiffle}).  The curves represent $E_\nu = 50,100,500,1000,5000,10^4\ \gev$ from bottom to top.  Fig. (b) is the differential cross section with parameters defined in Eq. (\ref{eq:nudiffhe}).  The curves represent incident neutrino energies
$E_\nu = 10^5,10^6,10^8,10^{10},10^{12}\ \gev$ from bottom to top.}
\label{fig:nudiff}
\end{figure}

Given our interest in high energies, we calculated the neutrino isoscalar nucleon cross section using the CTEQ6 parton distribution
functions \cite{cteq6} 
for $E_\nu = 50-10^{12}$ GeV \cite{gqrs,reno}. 
The numerical results use a small Bjorken $x$ extrapolation at very small $x$ according to a power law $xq(x,Q^2)\sim x^{-\lambda}$ \cite{gqrs,reno}.
We use the following parameterization for the differential cross section,
\begin{equation}
\label{eq:brdiff}
\frac{d\sigma_{cc}}{dy} = \frac{2m_pG_f^2E}{\pi}
\Bigl(a(E)+b(E)(1-y)^2\Bigr)\frac{1}{y^{c(E)}}\ .
\end{equation}
For neutrino scattering, we have two energy regimes split by neutrino energy
$E_c^\nu=3.5\times 10^4$ GeV, with 
\begin{eqnarray}
\label{eq:nudiffle}
a_\nu &=& 0.19-0.0265\,(2.214-\log (E_c/E))^2\\ \nonumber
b_\nu &=& 0.036 -0.0344\,(1.994-\log (E_c/E))^2\\ \nonumber
c_\nu &=& 2.3\times 10^{-2}\quad\quad E<E_c^\nu\ ,
\end{eqnarray} 
and for higher energies
\begin{eqnarray}
\label{eq:nudiffhe}
a_\nu &=& 0.060\,(E_c/E)^{0.675}\\ \nonumber
b_\nu &=& 0.169 \,(E_c/E)^{0.73}\\ \nonumber
c_\nu &=& 0.66 \times 10^p\\ \nonumber
p &=& 1.453(\log (E_c)/\log(E))^{6.24}
\quad\quad E>E_c^\nu\ .
\end{eqnarray}
Fig. \ref{fig:nudiff} compares the approximate form of the neutrino-nucleon differential cross section with Eq. (\ref{eq:brdiff}) and parameters from Eqs. (\ref{eq:nudiffle}-\ref{eq:nudiffhe}) with numerical results. 

\begin{figure}[h]
\includegraphics[width=2.4in,angle=270]{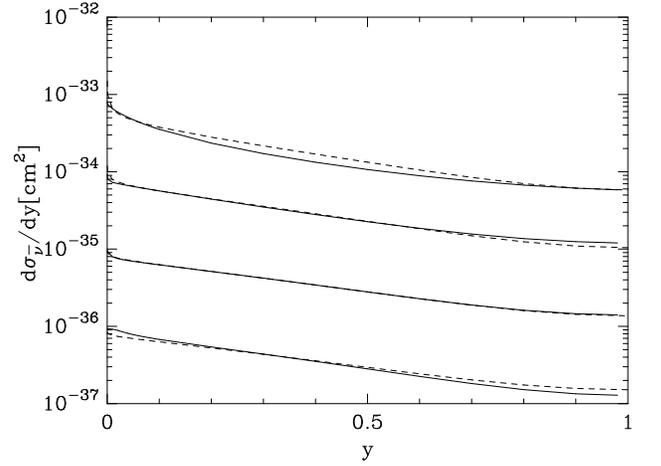}
\caption{Differential antineutrino-nucleon cross section defined by Eq. (\ref{eq:brdiff})
with parameters from Eq. (\ref{eq:anudiffle}). The solid lines represent numerical results using the CTEQ6 parton distribution functions \cite{cteq6} and the dashed lines are the results of our approximate analytic formula.  The curves represent incident antineutrino energies of $E_{\bar{\nu}} = 100,10^3,10^4,10^5\ \gev$ from bottom to top.}  
\label{fig:anudiff}
\end{figure}

Because the differential cross section is dominated by the contribution from sea quarks at high neutrino energy, we use the same high energy fit for antineutrino-nucleon scattering above $E_c^{\bar{\nu}}=10^6\ \gev$.  
The corresponding parameters for lower energy antineutrino-nucleon scattering are:
\begin{eqnarray}
\label{eq:anudiffle}
a_{\bar{\nu}} &=& 4.89\times 10^{-2} \times 10^{p_a}\\ \nonumber
p_a &=& -6.31\times 10^{-4}\log(E)^{4.05} \\ \nonumber
b_{\bar{\nu}} &=& 0.177\times 10^{p_b}\\ \nonumber
p_b &=& -2.78\times 10^{-5}\log(E)^{5.9} \\ \nonumber
c_{\bar{\nu}} &=& 4.4\times 10^{-3} E^{0.32}
\quad\quad E<E_c^{\bar{\nu}}\ .
\end{eqnarray}
Fig. \ref{fig:anudiff} shows the approximate form for antineutrinos for $E_{\bar{\nu}}\leq 10^5$ GeV with the parameters
from Eq. (\ref{eq:anudiffle}). 

The parameterizations for the $y$ distribution in neutrino and antineutrino scattering with isoscalar nucleons are within about 
15\% of the result using the CTEQ6 parton distribution functions \cite{cteq6}.

\section{Formalism for underground production of leptons}

\subsection{Electron production}

The probability for a muon to produce an electron via pair production as a function of muon energy ($E_\mu$) and emerging electron energy
($E_e^i$), in a depth interval $d\ell$, can be written as \cite{Dutta,Feng}
\begin{eqnarray}
P_{\rm prod}(\mu\rightarrow e,E_{\mu},E_e^i)=d\ell dE_e^i\frac{N_A}{A}\rho\frac{d\sigma_{pair}(E_\mu,E_e^i)}{dE_e^i}
\end{eqnarray}
where $N_A=6.022\times 10^{23}$ is Avogadro's number, $A$ is the atomic mass, and $\rho$ is the density of the material the muon is traversing.  For electrons produced in a detector (contained) 
which begins at depth $D$ and extends to
depth $ D+L_{\rm max}$. The event rate is given by 
\begin{eqnarray}
\label{eq:dndee}
\nonumber
\frac{dN}{dE_e^f} &=& \int _0^{L_{\rm max}} d\ell
\int_{E_e^i}^{\infty} dE_\mu \frac{dP_{\rm prod}}{d\ell}\phi_{\mu+\bar{\mu}}(E_\mu, D+\ell, \theta) \\ \nonumber
&\times& \int dE_e^i \delta(E_e^f-E_e^i)\\ \nonumber
&=& \int _0^{L_{\rm max}} d\ell
\int_{E_e^f}^{\infty} dE_\mu  \frac{N_A}{A} \rho \frac{d\sigma_{pair}(E_\mu,E_e^f)}{dE_e^f} \\ 
&\times & \phi_{\mu+\bar{\mu}}(E_\mu, D+\ell, \theta) \ , 
\end{eqnarray}
where we identify the initial electron energy $E_e^i$ with the electron
energy detected $E_e^f$. (This correspondence cannot be made for high energy taus.)
In the above equation, $\phi_\mu(E_\mu, D+\ell, \cos\theta)$ is the differential muon flux at depth $D+\ell$ as a function of energy and zenith angle $\theta$.

The same formalism can be applied to production of electrons via charged current electron neutrino interactions.  For production of single electrons or positrons from an incident neutrino or antineutrino the production probability has the same form with the replacement $d\sigma_{pair}(E_\mu,E_e^i)/dE_e^i\rightarrow d\sigma_{cc}(E_\nu,E_e^i)/dE_e^i$.  The incident differential muon flux at depth also needs to be replaced with an incident differential neutrino flux.

\subsection{Tau production}

Electrons and positrons must be produced in the underground detector to
be observed, however, very high energy taus can persist over long distances, although the tau loses energy in transit.
A tau with $E_\tau = 1$ PeV has a decay length of $\gamma c\tau \sim 50$ m.
The tau's electromagnetic energy loss over that distance is governed by
$\beta_\tau$ which is a factor of $m_\mu/m_\tau$ smaller than the corresponding energy loss parameter for muons. 
In the high energy limit ($E_\tau^i > 1\ \tev$) with continuous energy loss, 
\begin{equation}
\label{eq:eloss}
\langle \frac{dE}{dz}\rangle\simeq \frac{dE}{dz}\simeq -\beta_\tau E\ .
\end{equation}
The quantity $z$ is the column depth.
In the constant $\beta_\tau $ limit, the relation between tau initial energy
$E_\tau^i$ and its energy $E_\tau^f$ after traveling a distance $\ell$ in a material with a constant density
$\rho$ is
\begin{equation}
E_\tau^f = E_\tau^i \exp (-\beta_\tau\rho \ell)\ ,
\end{equation}
since $z=\rho \ell$.
We approximate  $\beta_\tau = 8.5\times 10^{-7}$ cm$^2$/g for standard rock \cite{drss,Dutta}.

The tau survival probability, accounting for tau energy loss and its finite lifetime, is \cite{drss,Dutta,Feng}
\begin{equation}
P_{\rm surv}(E_\tau^f,E_\tau^i) = \exp\Biggl[ \frac{m_\tau}{c\tau_\tau \beta_\tau \rho}
\Biggl( \frac{1}{E_\tau^i } -\frac{1}{E_\tau^f}\Biggr)\Biggr]\ .
\end{equation}
The short lifetime of the tau means that the survival probability goes to zero for low energy taus traversing any considerable distance.  Because of this, we will focus our calculation in the high energy limit. 

%%REV
We note that at high energies where $P_{\rm surv}\simeq 1$, a tau track
without a tau decay will mimic that of a muon with lower energy. The electromagnetic energy loss of the tau scales with energy according to
Eq. (12), so $\Delta E/\Delta z\simeq -\beta_\tau E_\tau$. Since $E_\tau$ is not know {\it a priori}, a muon with $E_\mu \simeq \beta_\tau E_\tau/\beta_\mu$ will show the same $\Delta E/\Delta z$. Taus that don't decay in a detector will be difficult to distinguish from lower energy muons.

The flux of taus produced from the atmospheric muon flux entering rock at sea level and which emerge
from rock of a total thickness $D$
is, including the survival probability of the tau, 
\begin{eqnarray}
\nonumber
\frac{dN}{dE_\tau^f} &=& \int_0^D d\ell \int_{E_\tau^i}^{E_\tau^i \exp(\beta_\tau\rho (D-\ell))}
dE_\tau^{i'} \int_{E_\tau^{i'}}^\infty dE_\mu \\ \nonumber
&\times & \phi_{\mu+\bar{\mu}}(E_\mu,\ell,\cos\theta) \frac{dP_{\rm prod}}{d\ell} \\ \nonumber
&\times& P_{\rm surv}(E_\tau^f,E_\tau^{i'}) \delta (E_\tau^f -E_\tau^{i'} \exp(-\beta\rho (D-\ell)) ) \\ \nonumber
&=& \int_0^D d\ell \int_{E_\tau^i}^\infty dE_\mu \frac{N_A}{A}\rho \frac{d\sigma_{pair}(E_\mu,E_\tau^i)}{dE_\tau^i} \\ \nonumber
&\times& \exp\Bigl[\frac{m_\tau}{c\tau\beta_\tau\rho}\Bigl(\frac{\exp(-\beta_\tau
\rho (D-\ell))}{E_\tau^f}-\frac{1}{E_\tau^f}\Bigr)\Bigl] \\
&\times& \phi_{\mu+\bar{\mu}}(E_\mu,\ell,\cos\theta) \ .
\end{eqnarray}
The delta function in the above equation explicitly enforces the energy loss relation found in Eq. (\ref{eq:eloss}).

Again, the same formalism for the production of tau pairs from an incident muon flux can be applied to the production of a single tau (antitau) particle from an incident tau neutrino (antineutrino) by replacing the incident muon flux with an incident tau neutrino flux. One also needs to replace the differential pair production cross section $d\sigma_{pair}/dE_\tau$ with the differential charged current neutrino cross section $d\sigma_{cc}/dE_\tau$, namely Eq. (\ref{eq:brdiff}).
 
\subsection{Atmospheric lepton fluxes}

Atmospheric lepton fluxes come from cosmic ray interactions in the atmosphere, producing mesons which decay to leptons \cite{Lipari1,bartol,honda,Bugaev,stasto,prs,tig,enberg}. 
For the differential muon flux from pion and kaon decay (the ``conventional flux'') at sea level, we use the analytical form from Ref. \cite{KBS} which can be written as a function of energy and zenith angle,
$\phi_{\mu+\bar{\mu}}(E_\mu,d,\theta)$, for depth $d=0$ at sea level:
\begin{eqnarray}
\label{eq:mukbs}
\phi_{\mu+\bar{\mu}}(E_\mu,0,\theta) &=& \frac{0.175 \ {\rm (GeV \, cm^2\, sr\, s)}^{-1}}  
{(E_\mu/\gev)^{2.72} }\\ \nonumber
&\times & \Bigl( 
\frac{1}{1+E_\mu\cos\theta^{**}/103\ \gev} \\ \nonumber
  &+& \frac{0.037}{1+E_\mu\cos\theta^{**}/810\ \gev} \Bigr).
\end{eqnarray}
Here $103\ \gev$ and $810\ \gev$ are the pion and kaon critical energies, respectively, which separate the high and low energy contributions to the atmospheric flux.  The effective cosine, $\cos\theta^{**}$, takes into account the spherical geometry of the atmosphere and is given by \cite{KBS}
\begin{eqnarray}
\cos\theta^{**} &=& S(\theta)\cos\theta^* \\
S(\theta) &=& 0.986+0.014\sec\theta .
\end{eqnarray}
The parameterization for $\cos\theta^*$ can be found in Appendix A of Ref. \cite{KBS}.

An additional contribution to the atmospheric flux comes from heavy flavor particle production and decay
in the atmosphere, the so-called prompt flux. For the prompt muon flux,
charmed meson production dominates. 
There are a number of predictions for the prompt muon flux \cite{tig,prs,enberg,pr,stasto}. 
The vertical prompt flux can be predicted from a perturbative QCD calculation which
can be parametrized by \cite{prs}
\begin{eqnarray}
\nonumber
\log_{10}(E_\mu^3 \phi_{\mu+\bar{\mu}}(E,0,\theta)) &\simeq& -5.37+0.0191 x
+0.156 x^2 \\
&-& 0.0153 x^3
\label{eq:muprs}
\end{eqnarray}
for $x=\log_{10}(E_\mu/\gev)$.
The results for atmospheric charm using a dipole model evaluation of the $c\bar{c}$ cross section
from Ref. \cite{enberg} give a lower prompt flux prediction. The approximate form for the sum of $\mu+\bar{\mu}$ at sea level 
from Ref. \cite{enberg} is 
\begin{equation}
\label{eq:muenberg}
\phi_{\mu+\bar{\mu}}(E_\mu,0,\theta) \simeq \frac{2.33\times 10^{-6}\, {\rm (GeV \, cm^2\, sr\, s)}^{-1}}{(E_\mu/\gev)^{2.53}
(1+E_\mu\cos\theta^{**}/E_0)}\ 
\end{equation}
where $E_0=3.08\times 10^6$ GeV. Both of the prompt flux formulas
are the same for the $\mu+\bar{\mu}$,
$\nu_\mu+\bar{\nu}_\mu$, and $\nu_e+\bar{\nu}_e$ fluxes, since to first approximation, the charmed mesons decay to electronic and muonic channels with equal branching fractions and the energy distribution of the muon and the muon neutrino is about the same.

The atmospheric muon flux at depth in rock or ice depends on electromagnetic energy losses of the muon as it 
passes through the material \cite{dedx,Lipari}.
Using the energy loss formula
\begin{eqnarray}
-\langle \frac{dE}{dz} \rangle = \alpha+\beta E\ ,
\end{eqnarray}
the muon flux at depth $d$ in a material of constant density $\rho$ can be written, to first approximation assuming continuous energy losses ({\it cl}), by
\begin{equation}
\label{eq:phidepth}
\phi_{\mu+\bar{\mu}}^{cl}(E_\mu, d, \theta) \simeq\phi_{\mu
+\bar{\mu}} (E_\mu^s, 0, \theta)
\exp (\beta\rho d)\ .
\end{equation}
Here, the muon energy $E_\mu$ at depth $d$ is related to the surface muon energy by
\begin{equation}
E_\mu^s = \exp(\beta \rho d) E_\mu +
(\exp(\beta \rho d) - 1)\alpha/\beta \ .
\end{equation}
The exponential factor in Eq. (\ref{eq:phidepth}) comes from
\begin{equation}
\frac{dE_\mu^s}{dE_\mu}=\exp(\beta\rho d)\ .
\end{equation}
The above formulas are valid for continuous energy losses.  Fluctuations in energy loss have the effect of increasing the down-going muon flux underground \cite{Lipari}, however, 
these corrections amount to 5-10\% enhancements for a depth of 1 km w.e. for muon energies between 100 GeV-1 TeV \cite{KBS}. For our evaluations here, we neglect the
corrections to the underground muon flux due to fluctuations in energy
loss.

\begin{figure}[h]
\includegraphics[width=2.4in,angle=270]{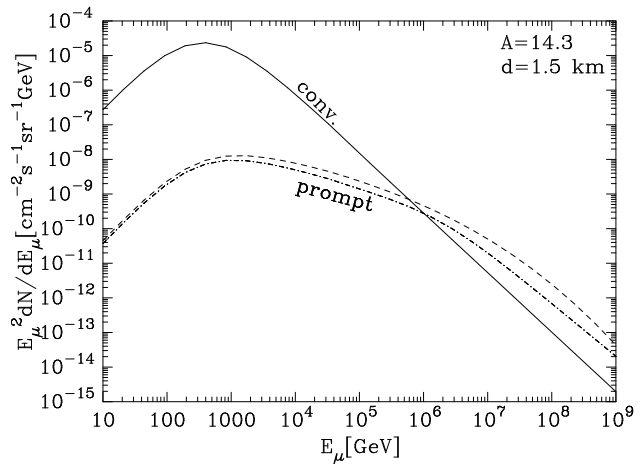}
\caption{Contributions to the underground muon flux from atmospheric conventional and prompt fluxes.  The solid line represents the contribution from the conventional atmospheric flux given in Eq. (\ref{eq:mukbs}).  The dashed curve represents the contribution from the atmospheric prompt flux given by Eq. (\ref{eq:muprs}) while the dot-dashed is for Eq. (\ref{eq:muenberg}).  These contributions are for a depth of $d = 1.5$ km in ice in the vertical direction.}
\label{fig:muflux}
\end{figure}

Fig. \ref{fig:muflux} compares the contributions to the underground muon flux at a depth of $d = 1.5$ km in ice ($A=14.3$). We use the two-slope parameterization of Ref. \cite{KBS} for the energy loss, with
\begin{eqnarray}
\label{eq:mubetalow}
\alpha & = & 2.67\times 10^{-3} {\rm GeV cm}^2/{\rm g}\\
\nonumber
\beta &=& 2.4\times 10^{-6} {\rm cm}^2/{\rm g}
\end{eqnarray}
for $E_\mu \leq 3.53\times 10^4$ GeV, and
\begin{eqnarray}
\label{eq:mubetahi}
\alpha & = & -6.5\times 10^{-3} {\rm GeV cm}^2/{\rm g}\\
\nonumber
\beta &= &3.66\times 10^{-6} {\rm cm}^2/{\rm g}
\end{eqnarray}
at higher energies.
We show the contributions from the vertical conventional flux given in Eq. (\ref{eq:mukbs}) as well as both of the prompt parameterizations given in Eqs. (\ref{eq:muprs}) (upper) and (\ref{eq:muenberg}) (lower prompt curve).  The prompt flux becomes the dominant contribution for muon energies $E_\mu \geq 10^6$ GeV.  This is due to the fact that charmed mesons decay more quickly than pions and kaons. The probability for pion and kaon decays introduces a factor of $1/E$ relative to the probability for
charm decays in the atmosphere for $E\lsim 10^8$ GeV.  

For the muon produced tau lepton pairs for a detector array
like HAWC, we need the atmospheric muon
flux at an altitude of 4.1 km. The high energy atmospheric 
muon flux at this altitude is approximately the same 
muon flux as at sea level. This is because the majority of the muons
are produced at an altitude of about 15 km \cite{pathlength}.
For the energies considered here, at altitudes between 15 km and 4 km,
pion and kaon energy loss through interactions with air nuclei are favored over meson decays.

Finally, to compare the electron and tau pair production rates from muons in transit to the rate for single electron and single tau production by
electron neutrinos and tau neutrinos respectively, we need the atmospheric electron and tau neutrino fluxes. At the energies considered here, $E>100$ GeV,
the conventional electron neutrino flux is approximately a factor of 135 smaller than the conventional muon flux \cite{bartol}. For our calculations here, we use
\begin{eqnarray}
\label{eq:nukbs}
\phi_{\nu_e+\bar{\nu}_e}(E_\mu,0,\theta) &=& \frac{1.30\times 10^{-3} \ {\rm (GeV \, cm^2\, sr\, s)}^{-1}}  
{(E_\mu/\gev)^{2.72} }\\ \nonumber
&\times & \Bigl( 
\frac{1}{1+E_\mu\cos\theta^{**}/103\ \gev} \\ \nonumber
  &+& \frac{0.037}{1+E_\mu\cos\theta^{**}/810\ \gev} \Bigr).
\end{eqnarray}
The conventional flux of electron neutrinos has an
approximate 60:40 ratio of $\nu_e:
\bar{\nu}_e$ at $E_\nu=1\ \tev$ \cite{bartol}, a ratio we use here for the full energy range.
For the prompt atmospheric $\nu_e+\bar{\nu}_e$ flux, we use
Eqs. (\ref{eq:muprs}) and (\ref{eq:muenberg}) as two representative
fluxes. The prompt neutrino to antineutrino ratio is 50:50.

The tau neutrino flux comes from two sources, oscillations from conventional neutrinos, primarily $\nu_\mu$, and from prompt decays of
the $D_s$ and $b$ mesons and subsequent tau decays \cite{pr,stasto}. The prompt tau neutrino flux from Ref. \cite{stasto} can be written approximately as
\begin{eqnarray}
\label{eq:nustasto}
\phi_{\nu_\tau+\bar{\nu}_\tau}(E,0,\theta) &=& \frac{1\times 10^{-7} E^{0.5}\ {\rm (GeV \, cm^2\, sr\, s)}^{-1}}{(E/\gev)^3} \\ \nonumber
&\times& \Bigl(\frac{1}{1+(E/1\times 10^6)^{0.7}+(E/4\times 10^6)^{1.5}}\Bigr) .
\end{eqnarray}
The average height of production of atmospheric leptons is at an altitude of $\sim 15$ km \cite{pathlength}. We are considering the tau neutrino downward flux or flux at $45^\circ$ zenith angle at energies above $\sim 50$ GeV.
The oscillation of $\nu_\mu\rightarrow\nu_\tau$ does not contribute significantly to the tau neutrino flux over these distances at the energies of interest, so we do not include it in our calculation.

\section{Results}

\subsection{Underground electrons in IceCube}
 
We begin with our results for the rate of electron production in large underground detectors.  For underground electrons, we have focused our calculation on large underground Cherenkov detectors such as IceCube.  We compare the flux of electrons produced via pair production by muons in transit through the detector with the flux produced by electron neutrinos or antineutrinos interacting via charged current interactions. We then show the contributions from muon bremsstrahlung which also produce an electromagnetic shower.
  
\begin{figure}[h]
\includegraphics[width=2.4in,angle=270]{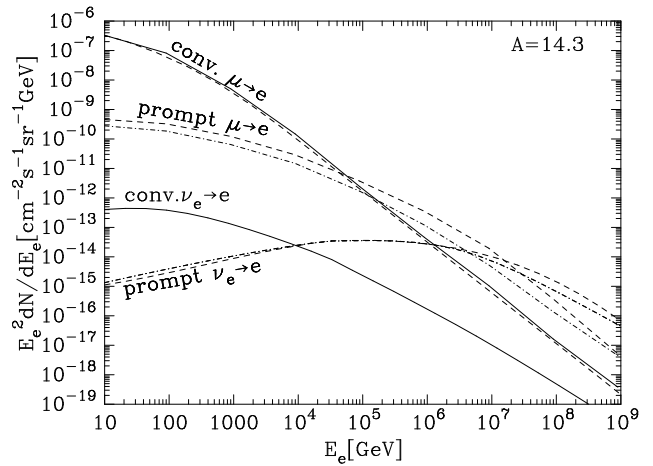}
\caption{The differential underground electron flux scaled by the square of the electron energy for electrons produced in ice between the vertical depths $1.5 \leq d \leq 2.5$ km.  The solid curves represent the electron flux produced by incident vertical conventional fluxes of muons and neutrinos given by Eqs. (\ref{eq:mukbs}) \& (\ref{eq:nukbs}).  The dashed curves labeled prompt represent the contribution from an initial prompt flux given by Eq. (\ref{eq:muprs}) while the dot-dash curve is from the prompt flux given in Eq. (\ref{eq:muenberg}). The dashed curve following the conventional
$\mu\to e$ curve was calculated using the Tannenbaum approximation to
$d\sigma_{pair}/dE_e$.}
\label{fig:fluxe}
\end{figure}

Fig. \ref{fig:fluxe} shows the differential flux of electrons and positrons as a function of electron energy produced in ice between the vertical depths of $1.5\ {\rm km}\leq d\leq 2.5\ {\rm km}$.  For muon induced events, we compare the contributions from incident conventional and prompt atmospheric muon fluxes, labeled $\mu\rightarrow e$ in the plot.  It is important to note that because the production mechanism is $\mu A\rightarrow\mu e^+ e^- X$, the total number of high energy events comes from the sum of the $\mu +\bar{\mu}$ atmospheric flux.  Accompanying the electron is a positron, a muon, and possibly evidence of the interaction with the target.  For comparison, we show the vertical conventional $\mu\rightarrow e$ rate using the Tannenbaum approximation to $d\sigma_{pair}/dE_e$ with the dashed line. 
 
We also show the contributions to the underground differential electron flux from an incident flux of electron neutrinos and anti-neutrinos, again
in the vertical direction.  For neutrino induced events, there is only one charged lepton produced, so the electron flux come purely from the incident neutrino flux while the positron flux comes from the incident antineutrino flux.  Because detectors like IceCube have no way of measuring charge, we have summed the rates electron and positron events together to show the total number of high energy events produced from incident neutrinos and antineutrinos to better compare with the number of events expected from atmospheric muon production.  
 
Roughly speaking, the conventional atmospheric muon flux falls approximately as $\sim E_\mu^{-4}$ at high energies and the prompt flux
falls as $\sim E_\mu^{-3}$. With the steeply falling muon flux, the production rate of electrons is dominated by the high energy tail of 
$d\sigma_{pair}/dE_e$. 
As can be seen in Fig. \ref{fig:fluxe}, the vertical underground electron flux is dominated by the conventional atmospheric muon flux for electron energies $E_e<10^4\ \gev$.  Around $E_e\simeq 10^5\ \gev$ the contributions from the prompt flux start to take over.  This is in contrast
to the crossover point for the muon flux, which occurs at an energy an order of magnitude higher, at $\sim 10^6$
GeV.  The electron energy distribution, with the electron accompanied by a muon, may augment efforts to measure the onset of the prompt muon flux.

For the downward electron neutrino and antineutrino fluxes, 
it is not until $E_e\simeq 10^7$ GeV that the contributions from incident neutrinos start to be comparable to those from incident muons.  The crossover
between the vertical conventional and prompt electron fluxes occurs at
$E_e\sim 10^4$ GeV. The electron production rate from
$\nu_e\rightarrow e$ is several orders of magnitude lower than from $\mu\rightarrow e$, except at the highest energies considered here.

Although our present analysis only considers atmospheric muon events in the vertical direction, it can be generalized to other zenith angles.
There are several features to consider with increased zenith angle. 
First,
the surface conventional atmospheric muon flux increases with angle, for example, by a factor $\sim 2$ for $E_\mu = 10^6$ when
the zenith angle increases from $0^\circ\rightarrow 60^\circ$. 
The surface convential atmospheric electron neutrino flux also increases with zenith angle.

At depth, muon and electron neutrino fluxes are affected differently.
Downward neutrinos experience little attenuation. Even at the highest energies, the neutrino interaction length from interactions with nucleons is larger than $10^5$ km of ice \cite{gqrs}. Muon energy loss from the surface to the detector is an important feature. The muon flux relative to the surface value reduces according to Eq. (22) in the limit of continuous energy loss. The net effect is that even though the surface
muon flux at non-zero zenith angles is larger than the vertical muon flux,
the flux for a detector at  depth $D$ (and slant depth $d\sim D/\cos\theta$) 
is decreased relative to the vertical muon flux at depth $D$. 
The underground flux of electrons produced in IceCube from an incident conventional atmospheric muon flux at a zenith angle of $30^\circ$ is reduced to $\sim 80\%$ of the flux produced from muons in the vertical direction for electron energy $E_e = 10$ GeV.  For electron energy $E_e = 10^9$ GeV, the electron flux produced by atmospheric muons with a zenith angle of $30^\circ$ is $\sim 90\%$ relative to the flux produced by vertical muons.  

\begin{figure}[h]
\includegraphics[width=2.4in,angle=270]{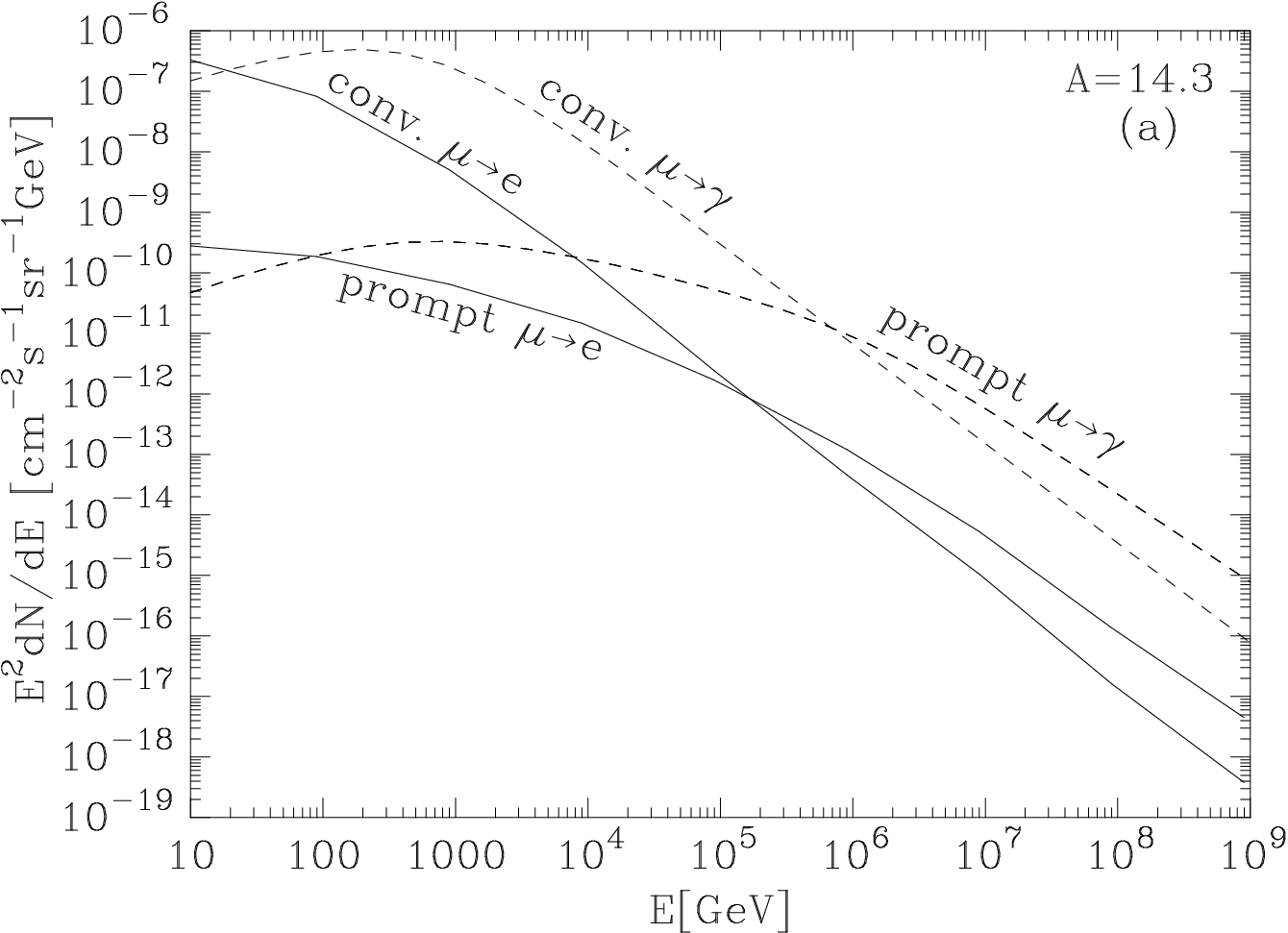}
\includegraphics[width=2.4in,angle=270]{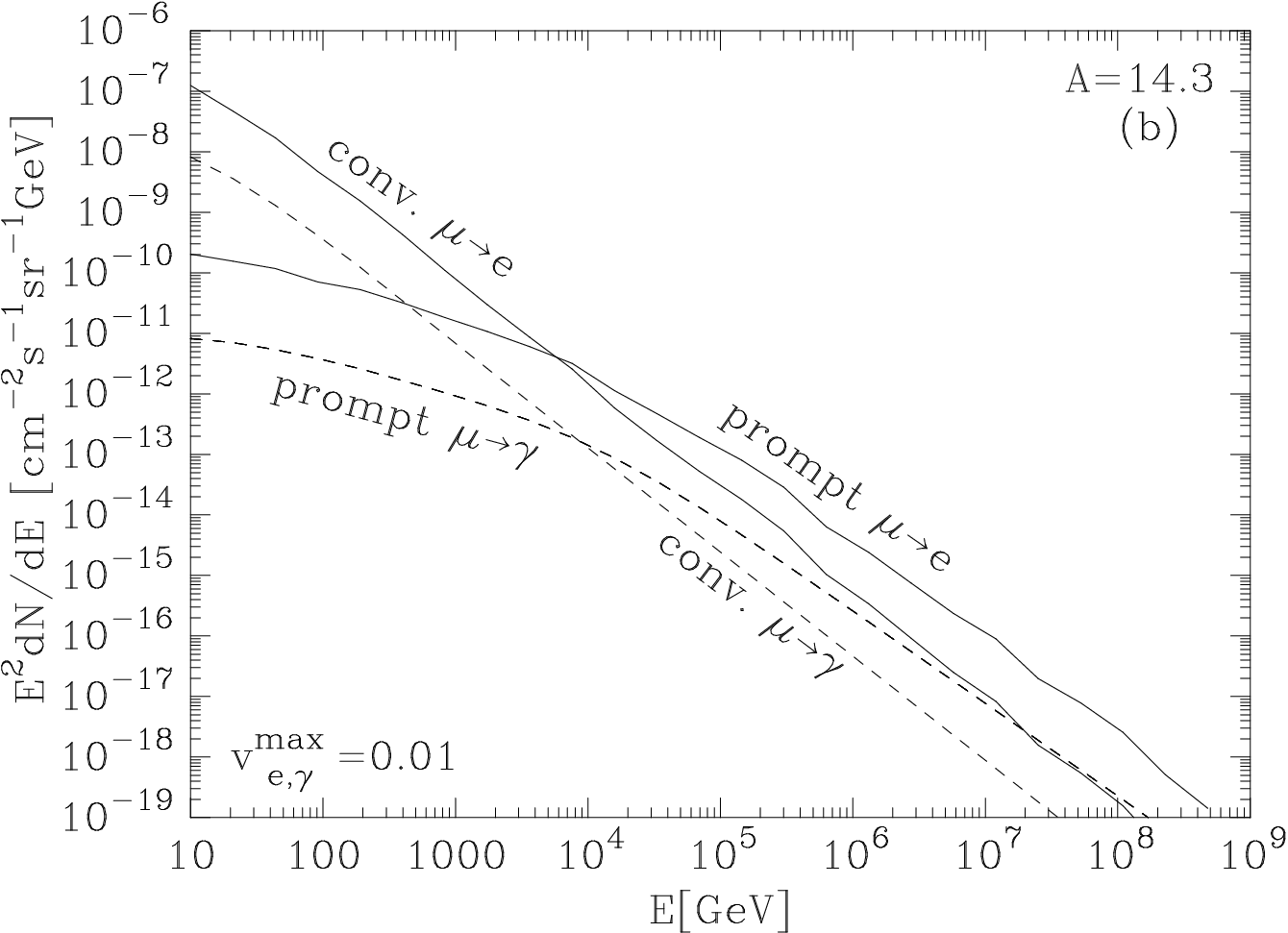}
\caption{The differential underground electron and photon fluxes scaled by the square of the electron or photon energy for particles produced in ice ($A=14.3$) between the vertical depths $1.5 \leq d \leq 2.5$ km.  Fig. (a) shows the total differential flux calculated using Eq. (\ref{eq:dndee}).  The solid curves represent the electron flux produced by incident vertical conventional and prompt (Enberg et al.) fluxes of muons. The dashed curves show the conventional and
prompt
$\mu\to \gamma$ contribution.  Fig. (b) represents the differential flux calculating by setting the upper bound on $v_{e,\gamma}$ to $v^{max}_{e,\gamma}=0.01$ ($E_\mu^{min}\approx 100E_{e,\gamma}$).  The curves are the same as in Fig. (a). }
\label{fig:fluxgamma}
\end{figure}
%%%%

At IceCube, the electromagnetic showers produced by photons look the same as the
electromagnetic showers produced by electrons. Muon bremsstrahlung in the
detector is therefore another source of electromagnetic showers.
In Fig. \ref{fig:fluxgamma}(a), we show with dashed lines the flux of photons
accompanied by a muon from $\mu A\rightarrow \mu\gamma X$ as a function of
photon energy. This flux is evaluated using the analytic formula of
Ref. \cite{brem} for the muon bremsstrahlung differential cross section. On the scale of the figure, the results are not much changed by including the more precise scaling of Ref. \cite{vanginneken}.

Above $E\simeq 20$ GeV, the bremsstrahlung contribution dominates the electromagnetic signal. This can be understood by the characteristic behavior of the energy distribution of photons versus electrons. The electron energy distribution from muon pair production falls more rapidly
with $E_e$ than $d\sigma_{\rm brem}/dE_\gamma\sim E_\gamma^{-1}$, so
when convoluted with the steeply falling atmospheric flux, the bremstrahlung contribution
dominates. The crossover between prompt and conventional photon signals moves to higher energies than the electron signals, but it is still less than the energy of the crossover of the muon flux itself.

Finally, we remark that the muon produced photon flux has a different relation between initial muon energy
and photon energy than the corresponding muon and electron energies.
This comes from the $E_\gamma$ scaling behavior which favors $v_\gamma =
E_\gamma/E_\mu> v_e=E_e/E_\mu$ for fixed $E_\mu$. If one could correlate
the incident $E_\mu$ to the outgoing $E_\gamma$ or $E_e$ and consider
$v_{e(\gamma)}<v^{max}_{e(\gamma)}=0.01$, the electron flux from conventional muons would dominate the the photon flux from conventional muons. The differential flux of photons and electrons with this scaling restriction is shown in Fig. \ref{fig:fluxgamma}(b).  For 
such a restriction in $v_{e(\gamma)}$, the electron flux from conventional muons drops by a factor of $\sim 10^2$, however, the photon
flux from conventional muons would be reduced by a factor of $\sim 10^5$.
The electron signal dominates the bremsstrahlung signal by about a factor
of 10 with this restriction. Electromagnetic showers from muons still dominate the electron neutrino induced electromagnetic showers. 

\subsection{PeV taus at IceCube}

Muon production of high energy electrons and positrons may be a
background to searches for taus via ``lollipop'' events \cite{cowen}. 
At these energies, muons do not decay, but taus do, yielding an event with a ``muon-like'' track which has a splash of energy
from the tau decay. The tau track, before the decay, appears as a muon-like track.
The high energy electron from $\mu A\to \mu e^+ e^- X$
could also leave an energy splash, with a continuing muon as part of the
event configuration. Furthermore, there is the potential to produce 
tau pairs by $\mu A\to \mu \tau^+ \tau^- X$. Some of these taus will appear at the edge of the detector, and some of them will decay in the detector. 

\begin{figure}[h]
\includegraphics[width=2.4in,angle=270]{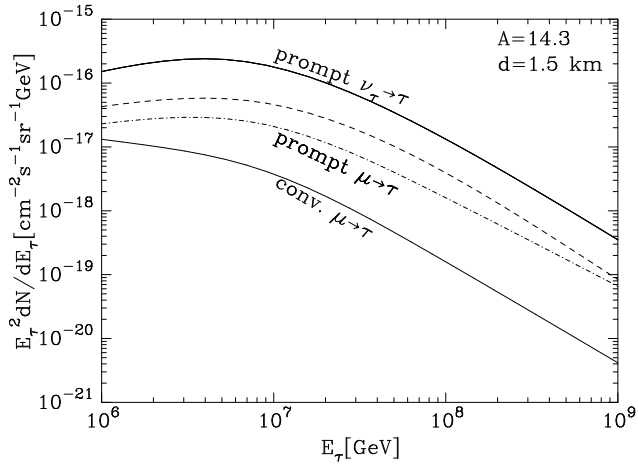}
\caption{Differential tau flux scaled by the square of the final tau energy entering the detector at a depth of $1.5$ km in ice.  The lower solid line corresponds to tau production from a vertical incident conventional muon flux given in Eq. (\ref{eq:mukbs}).  The dashed and dot-dashed curve represents the tau flux from a vertical incident prompt flux given in Eq. (\ref{eq:muprs}) and Eq. (\ref{eq:muenberg}) respectively.  The top solid curve is for the tau flux produced with an incident prompt tau neutrino flux given in Eq. (\ref{eq:nustasto}).}
\label{fig:fluxtauice}
\end{figure}

To explore tau production by atmospheric muons and atmospheric tau neutrinos and antineutrinos, we have calculated the vertical tau flux entering the IceCube detector at a depth of $1.5$ km in ice for tau energies between $1-1000$ PeV.  Fig. \ref{fig:fluxtauice} shows the different contributions to the tau flux from incident atmospheric lepton fluxes.  At a depth of $1.5$ km, the tau events are dominated by the charged current production from the incident prompt tau neutrino flux given in Eq. (\ref{eq:nustasto}).  In this energy range, the contribution to the underground tau flux from an incident atmospheric muon comes mainly from the prompt flux. 

To convert to lollipop events, one needs to multiply the tau flux by the decay probability. Since the decay length at $10^6$ GeV is about 50 m,
the decay probability over the 1 km of the detector at that energy is $P_{decay}=1$, and it decreases to
$P_{decay}\simeq 0.2$ at $E_\tau = 10^8$ GeV. The flux of decaying taus, where the taus are produced by muons or tau neutrinos, is quite small
compared to the fluxes shown in Fig. \ref{fig:fluxe}.

While we have not done a full scale analysis of the number of events with decaying taus in the detectors, we can use a characteristic
\begin{equation}
\label{eq:isotropice2}
\phi_{\nu+\bar{\nu}}=10^{-8}\, ({\rm cm^2s\,sr\,GeV})^{-1}({\rm GeV}/E)^2
\end{equation}
isotropic neutrino flux and look at the relative normalizations. At $E_\nu=10^5$ GeV, Eq. (\ref{eq:isotropice2}) gives a flux that is a factor of about 30 larger than the prompt $\nu_\tau$ flux used in Fig.
\ref{fig:fluxtauice}. The downward tau flux produced near the edge of the detector from this $E^{-2}$ tau neutrino flux would remain at least a factor of 30 larger than the prompt $\nu_\tau\to\tau$ contribution shown.
It is clear that the prompt tau neutrino flux will be quite difficult to see in IceCube, as will be the high energy tau flux from muons in transit.

\subsection{Tau production for HAWC}

A different geometry for muon production of taus is to use a mountain as the muon conversion volume.  For sufficiently high energies of the produced taus, the taus can exit the mountainside and decay in the air.
The proposed HAWC surface array has the potential
to measure the tau decay air shower.  
This detector sits in a mountain saddle at an altitude of $4.1$ km shielded by mountains on two sides.  

\begin{figure}[h]
\includegraphics[width=2.4in,angle=270]{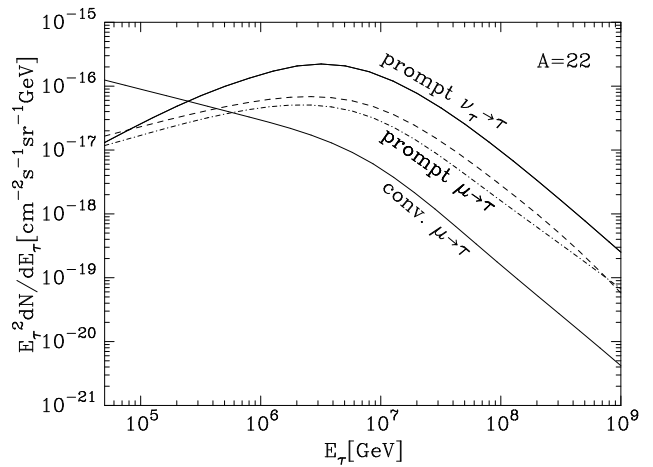}
\caption{Differential tau flux scaled by the square of the final tau energy emerging from $1$ km water equivalent of rock.  We use a zenith angle of $45^\circ$ for our incident fluxes.  The lower solid line corresponds to tau production from an incident conventional muon flux given in Eq. (\ref{eq:mukbs}).  The dashed and dot-dashed curve represents the tau flux from an incident prompt flux given in Eq. (\ref{eq:muprs}) and Eq. (\ref{eq:muenberg}) respectively.  The top solid curve is for the tau flux produced with an incident prompt tau neutrino flux given in Eq. (\ref{eq:nustasto}).}
\label{fig:fluxtau}
\end{figure}

For our calculation, we have used a zenith angle of $45^\circ$ in the incident flux and $1$ km water equivalent distance of rock for the incident muons or tau
neutrinos in transit.  
Fig. \ref{fig:fluxtau} shows the energy distribution of taus emerging from the rock for both incident muon and prompt tau neutrino atmospheric fluxes.  We also show the contributions from an incident prompt muon atmospheric flux.  

As can be seen in the plot, at energies of $E_\tau < 10^5$ GeV the dominant contribution to the emerging tau flux comes from the conventional atmospheric muon flux.  As noted above, a tau energy of $E_\tau =10^5$ GeV has a decay length of $\sim 5$ m.  Even at $E_\tau = 10^7$ GeV,
the decay length of 500 m may allow the tau decay to be measured by
HAWC. At energies above $E_\tau \simeq 10^6$ GeV,
the contribution of taus produced by atmospheric muons provides about a $20-30\%$ contribution to the total ``atmospheric'' tau flux which is dominated by atmospheric tau neutrino conversions. As is the case with taus in IceCube from atmospheric sources, the event rates in HAWC would be quite small.

\section{Discussion}

Our evaluation of the pair production cross section in Ref. \cite{br} already showed that Tannenbaum's parameterization of the cross section \cite{br} is a valuable shortcut. Here, we have also shown that his differential distribution for electron positron pair production, with the identification of $v=(E_\mu-E_\mu')/E_\mu\simeq 2 E_e/(E_\mu-E_e/2)$, is a reasonably good representation of the numerical evaluation of the electron energy distribution.  Using the approximate form from Tannenbaum for the differential cross section yields errors that are $\leq 30\%$ in the calculation of underground electron fluxes for electron energies $10\ \gev\leq E_e \leq 10^9\ \gev$.

Underground electron and photon production by atmospheric muons may aid in our understanding of the atmospheric lepton flux itself. The crossover point for prompt versus conventional sources in terms of the electron energy distribution is at a lower energy
than the crossover point for muons. Comparing Figs. \ref{fig:muflux} and \ref{fig:fluxe} shows that the crossover happens one order of magnitude lower when looking at pair produced electrons as opposed to muons for a vertical depth of $\sim 1.5$ km.  Fig. \ref{fig:fluxgamma} shows that for
the full bremsstrahlung signal, the crossover point is intermediate between
electrons and the incident muons.
The additional information gained in studying underground electromagnetic
 signals may aid in the determination of the correct charm production model.

The energy threshold for ``lollipop'' events from tau decays in IceCube is $\sim 5$ PeV.  A $5$ PeV tau will produce $>200$ m track length in the detector \cite{cowen}.  Although potentially difficult to see, the prompt $\nu_\tau\rightarrow \tau$ flux entering the detector receives $\sim 20\%$ contribution from high energy taus produced by prompt muons in transit through ice. The conventional $\mu\rightarrow\tau$ flux is suppressed by about two orders of magnitude in this energy regime relative to prompt $\nu_\tau\rightarrow\tau$ flux. 
 
When calculating the underground electron or positron fluxes from pair production, there should be rare events where one of the leptons comes out with a significant fraction of the initial muon energy.  In a detector like IceCube, it may be difficult to identify the accompanying lepton, as well as the muon after scattering.  This type of event could fake a ``lollipop'' type signal when looking for tau neutrino induced tau events. A comparison of Figs. 6 and 7 shows that the flux of electrons at an energy of a few PeV is about a factor of $\sim 100$ times the flux of atmospheric $\nu_\tau$ induced taus. Muon bremstrahlung contributions are even larger, potentially adding to a faked signal.

While muon production of tau pairs in the PeV energy range is less than the prompt atmospheric $\nu_\tau$ contribution,
at lower energies there is the potential for downward secondary $\nu_\tau$ production from conventional muons via
\begin{eqnarray}
\mu A&\rightarrow& \mu\tau^+\tau^- X \\ \nonumber
\tau&\rightarrow& \nu_\tau X\ ,
\end{eqnarray}
in, for example, the TeV to PeV energy range.
In this energy range, the taus decay promptly.

Our calculations have focused on large Cherenkov detectors, but the same formalism could be applied for calculating underground lepton rates at the Indian Neutrino Observatory (INO).  Due to magnetization, INO will have the capability of separating the electron (or $\tau^-$) signal from that of the positron (or $\tau^+$).  This will allow observations at INO to determine the energy distribution of the charged partner when looking at pair production events, something that could be predicted numerically with our evaluation of the differential cross section.  

Electromagnetic interactions, in particular lepton pair production by muons in transit through materials, are interesting in their own right, not just how they affect the energy loss of muons. As neutrino telescopes and air shower detectors focus on neutrino induced signals, muon signals with high energy electrons or taus may provide interesting cross checks to neutrino signals.

%============================================================================== 
%============================================================================== 
\begin{acknowledgments}
This research was supported by 
US Department of Energy 
contract DE-FG02-91ER40664. We thank T. DeYoung for helpful conversations.

\end{acknowledgments}

%=============================================================================

\end{document}